\documentclass[12pt]{JHEP3}
\usepackage{epsfig}

\renewcommand{\)}{\right )}
\renewcommand{\[}{\left [}
\renewcommand{\]}{\right ]}

\def\fslash#1{#1 \!\!\! \slash}
\def\beq{\begin{equation}}
\def\eeq{\end{equation}}
\def\pa{\partial}

\def\varp{\varepsilon}
\def\bea{\arraycolsep .1em \begin{eqnarray}}
\def\eea{\end{eqnarray}}

\def\vk{{\bf k}}

\def\Tr{{\rm Tr}}
\def\rnn{{\rho NN}}
\def\rpp{{\rho \pi\pi}}
\def\grnn{{g_{\rho NN}}}
\def\grnns{{g_{\rho NN}^2}}
\def\mss{{M_N^*}^2}
\def\kr{{\kappa _\rho}}
\let\de=\delta

\let\om=\omega

\let\no=\nonumber

\def\eq#1{(\ref{#1})}
\def\refr#1{\cite{#1}}
\def\refrs#1{Refs.\cite{#1}}

\def\s0#1#2{\mbox{\small{$ \frac{#1}{#2} $}}}
\def\0#1#2{\frac{#1}{#2}}

\title{{\bf Spectral Function of $\rho $ in Dense and Hot Hadronic
Matter
}
}
\author{Ji-sheng Chen$^{a,b}$~~~~~~
Jia-rong Li$^{b}$~~~~~~Peng-fei Zhuang$^{a}$\\
$^a$Physics Department, Tsinghua University, Beijing 100084, P.R.
China\\
$^b$Institute of Particle Physics,
 Hua-Zhong Normal University,
Wuhan 430079, P.R.China.\\
Email: \email{jschen@mail.tsinghua.edu.cn}}
\abstract{
The spectral function of rho meson in hot/dense hadronic matter is
studied by taking into account the nucleon-loop on quantum
hadrondynamics model level.
Different from the hot pion gas effect which changes the spectral function
slightly, the nucleon-antinucleon polarization
(Dirac sea) makes the spectral 
function very sharp and shifted towards the low invariant mass 
region significantly due to the decreasing effective nucleon
mass.
}\vskip 0.15cm
\preprint{.}
\begin{document}
\par 
It is widely prospected that in
ultra-relativistic heavy-ion collisions a new phase of matter---quark-gluon plasma (QGP) may be generated
and the spontaneously broken chiral symmetry
may be restored. Among the proposed various signals
which can detect the phase transitions in strongly 
interacting matter, the electromagnetic signals-dileptons and photons are
considered to be the clearest ones, since they can penetrate the medium
almost undisturbed. Considering the fact that light vector mesons can directly decay
to dilepton pairs, the study about $\rho $ meson in medium
is especially interesting because the lifetime of $\rho$ is smaller than
that of the fireball formed in heavy-ion collisions and the decay of $\rho $ can be
completed inside it. The strong enhancement of
low mass dileptons in  central $A-A$ collisions
observed by CERES-NA45 has excited many theoretical
works\refr{agakichiev,wurm}. Among them, the
mass dropping of $\rho $ mesons provides
a good description of the data\refr{brown,rapp,li,hatsuda}.
\par
The mass spectrum of hadrons in extreme hadronic
environment has been widely discussed.  
One of the familiar results is the Brown-Rho scaling law indicating that the masses of nucleons and mesons decrease
in hot/dense hadronic environment. For light vector mesons, it is difficult to
extract an unique conclusion from different model calculations. 
The mean field theory (MFT) and effective
Lagrangians can give similar results as the Brown-Rho scaling
\refr{rapp,saito}, but the results from some QCD sum rules may be different from it and even
contrast to it\cite{leupold,klingl,hatsuda1}. The property of $\rho $ in
hot/dense environment is not yet clear and needs further study.
 
The hot topic in recent years is discussing the low
invariant mass dilepton production in heavy ion collisions. 
The dilepton distribution is related with the spectral function of
$\rho $ meson, i.e., the imaginary part of the retarded
propagator of $\rho $ in medium \refr{kapusta,weldon,rapp2}, which reflects the comprehensive medium
effects.   From the point of view of the partial chiral symmetry
restoration the $\rho $ spectral function has been widely discussed in recent
literature such as in \refrs{rapp,rapp2} (and references therein).

The key point of discussing the
$\rho $ spectral function is how to calculate the $\rho $ meson self-energy in nuclear
environment. In this paper, the effects of the decreasing effective 
nucleon mass on the $\rho $ spectral function
in medium are discussed.
With quantum hadrodynamics model(QHD) of nuclear matter,
the effective nucleon mass drops down with density/temperature and 
should be taken into account in discussing the effective mass as well as the decay
width  of $\rho $ in medium, i.e.,
the spectral function.
H. Shiomi and T. Hatsuda have found that the effective $\rho $
meson mass decreases in nuclear matter due to the nucleon polarization 
at finite density with QHD-I type effective
Lagrangian\cite{shiomi}. By extending their work
to finite temperature, the temperature and
density dependence of the $\rho $-meson spectral function due to nucleon loop is discussed.

\par 
 In Minkowski space, the
self-energy of the light vector meson $\rho $ can be expressed as\refr{kapusta,kapusta1}
\bea \Pi ^{\mu \nu } = \Pi _L P_L^{\mu\nu}
+\Pi _T P^{\mu \nu }_T,
\eea 
where $k^2=k_0^2-\vk ^2$. The $P^{\mu\nu }_L(k) $ and
$P^{\mu\nu }_T (k)$ are the standard longitudinal and transverse
projection tensors
\bea\label{tensor}
P_T^{00}&&=P_T^{0i}=P_T^{i0}=0,~~~
P_T^{ij}=\de ^{ij}-\0{k^ik^j}{\vk ^2},\no\\
P_L^{\mu\nu}&&=\0{k^\mu k^\nu}{k ^2}-g^{\mu\nu}-P_T^{\mu\nu}.
\eea
The longitudinal and transverse elements of the
self-energy can be determined by
\bea
\Pi_L(k)=\0{k^2}{\vk ^2 }\Pi ^{00}(k),~~~~~
\Pi_T(k)=\012 P^{ij}_T\Pi_{ij}(k).
\eea
Noting that the
self-energy is related to the full ($\cal D$) and bare
(${\cal D}_0$) propagators by 
\bea \Pi ^{\mu\nu }=({\cal
D}^{-1})^{\mu\nu}-({\cal D}^{-1}_0)^{\mu\nu}, 
\eea 
one can obtain 
\bea {\cal D}^{\mu\nu }=-\0{P^{\mu\nu }_L}{k^2-m^2_\rho
-\Pi _L}-\0{P^{\mu\nu }_T}{k^2 -m_\rho ^2 -\Pi_T}-\0{k^\mu k^\nu
}{m^2_\rho k^2 }, 
\eea 
where $m_\rho $ is the free mass of $\rho$ in vacuum.

The property of in-medium $\rho $ is determined by its full propagator.
The imaginary part of the retarded propagator is referred to as the spectral
function and related to the dilepton production with vector meson dominance
model. The study of the $\rho $ spectral function is attributed
to calculating the in-medium self-energy.
By concentrating on decreasing nucleon mass effects,  
the in-medium $\rho $ property is analysed in terms of the Feynman diagrams
displayed in Fig.\ref{fig1}.
\FIGURE{
\psfig{file=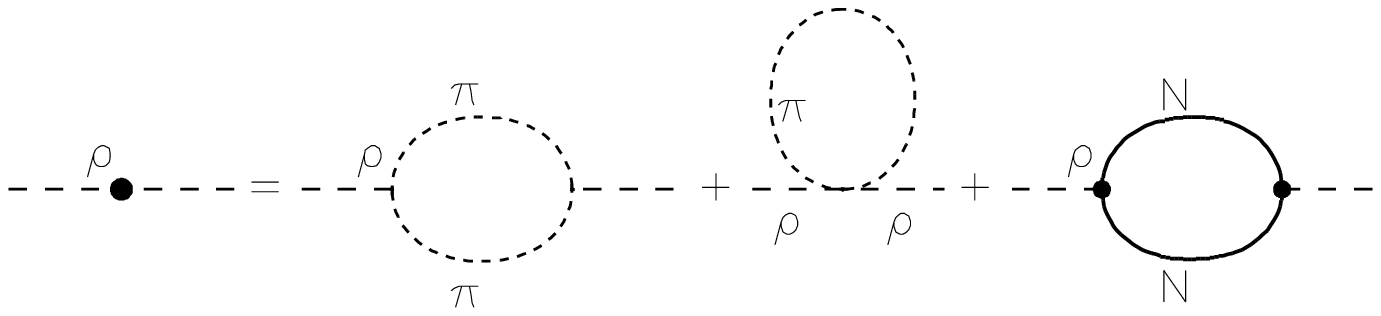,width=10 cm,angle=-0}
\caption{
\small The $\rho $ self-energy Feynman diagrams associated with
$\rpp$ and $\rho \rho \pi \pi $ as well as $\rnn$ interactions. }\label{fig1}
}
The pion effect has been analyzed in Ref.\cite{kapusta}, and in Minkowski space
the pion-loop contribution to $\rho$ self-energy can be written as
\begin{equation}\label{one} 
\Pi _{\mu\nu}^{\rpp}(k)=-g_\rho ^2 T
\sum _{p_0}\int \frac{d^3 p}{(2 \pi)^3} \frac{(2 p+k)_\mu(2
p+k)_\nu}{(p^2-m_\pi ^2)[(p+k)^2-m_\pi^2 ]}+2 g_{\mu\nu} g^2_\rho
T\sum _{p_0}\int \frac{d^3p }{(2 \pi )^3}\frac{1}{p^2-m_\pi ^2},
\end{equation}
where the 0-component of $\pi$ loop momentum is related to temperature via 
$p_0= 2 n \pi T i$, with $n=0,\pm 1, \pm 2, \cdots $. 
\par
The contribution of nucleon excitations through nucleon-loop to $\rho$ self-energy
 is analyzed in terms of the
effective Lagrangian\cite{shiomi,physrep}
 \bea {\cal L}_{\rho NN }=g_{\rho
NN} ({\bar \Psi} \gamma _\mu \tau ^a \Psi V_a ^\mu - \0{\kappa
_\rho}{2 M_N} {\bar \Psi}\sigma _{\mu\nu }\tau ^a\Psi \pa ^{\nu
}V_a^\mu ), \eea where $V_a^\mu$
is the $\rho$ meson field and $\Psi $ the nucleon field.
It can be written as
\bea\label{rnns} \Pi _{\mu\nu}^{\rnn}(k)=2 \grnns
T\sum _{p_0} \int \frac{d^3 p}{(2 \pi)^3} \Tr
\[\Gamma _\mu (k) \frac{1}{\fslash{p} -M_N^*}\Gamma _\nu (-k)
\frac{1}{(\fslash{p}-\fslash{k}) -M_N^*}\],
 \eea
where 
$\Gamma _\mu $
is defined as 
$
\Gamma _\mu (k) = \gamma _\mu +\0{i  \kr }{2 M_N}\sigma
_{\mu\nu }k^\nu $
with $\sigma _{\mu\nu }=\0{i[\gamma _\mu, \gamma
_\nu]}{ 2}$, $M_N$ and $M^*_N$ are the nucleon masses in vacuum and medium, respectively. 
The 0-component of nucleon loop momentum is related to temperature
via $p_0=(2 n +1 ) \pi T i+\mu ^*$ with $\mu ^*$
being the effective chemical potential of nucleon.

\par
With residue theorem the $\rnn $ contribution
can be separated into two parts
\bea\label{seven}
\Pi _{\mu\nu} ^{\rnn}(k)=(\0{k_\mu k_\nu}{k^2}-g_{\mu\nu}) \Pi
_F^{\rnn}(k)+\Pi _{D,\mu\nu} ^{\rnn}(k). 
\eea
 The first
part $\Pi
_F^{\rnn}(k)$ corresponding to $\Pi ^F$ of Ref. \cite{shiomi} 
contains divergent integrals and needs to be regularized. However, it can not be
renormalized by the conventional method. Using the
dimensional regularization method and taking a phenomenological procedure
\cite{shiomi,peskin,Kurasawa,mazumder}  
\bea
\pa ^n \Pi ^\rnn _F (k) /\pa
(k^2)^n|_{M^*_N\rightarrow M_N, k^2=m^2_\rho }=0,~~~~ ( ~~n=0,1,2,\cdots, \infty
~~ )\eea
one can eliminate the real vacuum ($T=0$) part and obtain the finite result
\bea\label{rnn1} 
\Pi_{F,L(T)}^{\rho NN} (k) =&&k^2
\0{g_{\rho NN}^2 }{\pi ^2 } \(P_1 +\0{\kr M_N^*}{2 M_N} P_2
+ (\0{\kr}{2 M_N})^2\0{k^2 P_1+{M^*_N}^2P_2}{2}\),\\ 
 P_1=&&\int ^1_0 dx x
(1-x) \ln c
,~~~~~~
P_2 =\int _0^1 dx \ln c
,\no\\
c=&&\0{{M_N^*}^2 -x(1-x)
k^2}{{M_N}^2 -x(1-x) k^2}.\no\eea
It is easy to see that the medium effect on  $\Pi _F^\rnn $ is also introduced by
 the effective nucleon mass $M_N^*$ in hadronic
environment which will be determined by QHD-I below.

\par
The second part $\Pi _{D,\mu\nu} ^{\rnn}(k)$ in \eq{seven}
is explicitly related to distribution function $n_N(T,\mu
^*)$(corresponding to the Fermi sea contribution at $T=0$).
Its calculation is lengthy. We list here only the longitudinal results
\bea\label{rnn2}
\Pi_{D,L}^{\rho NN}(k)=&&\Pi_{1D,L}^{\rho
NN}(k)+\Pi_{2D,L}^{\rho NN}(k)+\Pi_{3D,L}^{\rho
NN}(k),\\
\Pi_{1D,L}^{\rho NN}(k)=&&-\0{\grnns k^2}{\pi ^2\vk ^2
}\int \0{ p^2 dp}{\om } (n_N +{\bar n}_N) \[2\right.\no\\
&&~~~\left.+\0{k^2 - 4 \om
k_0 +4 \om ^2 }{4 p |\vk |} \ln a +\0{k^2 +4 k_0 \om  +4 \om ^2
}{4 p |\vk |} \ln b
\],\no\\
 \Pi_{2D,L}^{\rho
NN}(k)=&& \0{\grnns k^2 }{\pi ^2|\vk |
} \0{\kr }{2 M_N} M_N^* \int \0{p dp}{\om} (n_N+{\bar n }_N)\[
\ln a +\ln b
\],\no\\
\Pi_{3D,L}^{\rho
NN}(k)=&& \0{\grnns k^2 }{\pi ^2 \vk ^2 }(\0{\kr }{2 M_N})^2\int
\0{p^2 dp }{\om } (n_N +{\bar n}_N)\[2 k_0^2 \right.\no\\
&&~~~\left.+\0{\vk ^2 (k^2 -4 p^2)
+(k^2 -2 \om k_0 )^2 }{4 p |\vk |}\ln a+\0{\vk^2 (k^2 -4 p^2) +(k^2 +2 \om k_0 )^2 }{4 p |\vk |} \ln b
\], \no\eea
where $a$, $b$, $\om $,  $n_N $,  ${\bar n_N} $
are defined by 
\bea a&&= \0{k^2 - 2 p |\vk |- 2 k_0 \om }
                {k^2 + 2 p |\vk |- 2 k_0 \om },
~~~~~~
b= \0{k^2 - 2 p |\vk |+ 2 k_0 \om } {k^2 + 2 p |\vk
|+ 2 k_0 \om }, \\
\om &&=\sqrt{p^2+\mss },~~~~~ n_N=\0{1}{e^{\beta (\om
 -\mu ^* )}+1},~~~~~ {\bar n}_N=\0{1}{e^{\beta (\om +\mu ^* )}+1}\no
.\eea

Before making further discussions, 
it should be emphasized that the inclusion of the pion tadpole diagram is
very crucial for the current conservation condition(gauge
invariance)\refr{kapusta}.
It is interesting that the self-energy is divided into two parts
such as in \eq{seven} with residue theorem and the two parts separately satisfy the current
conservation. The
current conservation condition ensures the transversality of $\Pi ^{\mu\nu}$,
which is reflected in the properties of the projection tensors $P_L^{\mu\nu}$
and $P_T^{\mu\nu}$ such as $k_\mu P_{L(T)}^{\mu\nu}=0$.
\par

After analytical continuation according to $k_0\rightarrow E+i \varp $,
$E=\sqrt{M_\rho ^2+{\bf k}^2}$ with $M_\rho $
being the $\rho $ invariant mass in medium, one can obtain the real and
imaginary parts of the self-energy.
Then by using the definition of $\rho$ meson spectral function
\bea\label{ten}{\cal A}
_{L(T)} (k)=- 2\0{Im \Pi  _{ L(T)} (k)}{\[M_\rho^2 -
(m_\rho ^2 +Re\Pi ^\rho_{ L(T)} (k ))\]^2+\[Im\Pi
^\rho_{L(T)} (k ) \]^2 },\eea 
where $\Pi _{L(T)}$ is the total longitudinal or transverse self-energy,
one can analyze the medium effects on $\rho $ meson.

The effective mass $M_N^*$  and effective chemical potential $\mu ^*$ of the nucleon
in $\Pi ^{\rho NN}$ are determined from the simplest QHD-I at mean field
level\refr{serot1986,serot,zhang}.
 The QHD-I is a renormalizable model including nucleon and meson degrees of freedom
\bea
{\cal L}=&&{\bar \psi }\[\gamma _\mu (i \pa ^\mu -g_v V^\mu )-(M_N-g_s
\phi )
\]\psi +\012 (\pa _\mu \pa ^\mu \phi -m_s^2 \phi ^2)\no\\
&&~~~~~~~~~~~~~~~~~-\014 F_{\mu\nu
}F^{\mu\nu } +\012 m_v^2V_\mu V^\mu +\delta {\cal L},
\eea
where $\psi $ is the nucleon field, $ \phi $ the
neutral scalar field with free mass $m_s$,
$F_{\mu\nu}=\pa _\mu
V_\nu -\pa _\nu V_\mu$ the field tensor of $\om $ meson  with free mass
$m_v$, and $\delta
{\cal L}$ the renormalization counterterm.

In mean field approximation, the meson field operators are replaced by the
ground-state expectation values
\bea
\phi &&\rightarrow <\phi >=\phi _0,~~~~~
V^\mu \rightarrow <V^\mu >=g^{\mu 0} V^0.
\eea
With MFT by neglecting the vacuum fluctuation contribution, the classical field $\phi _0$ is a dynamical quantity which makes
the nucleon mass drop and can be determined self-consistently
by\refr{serot1986,zhang}
\bea
M_N-M_N^*=g_s\phi _0 =g_s ^2 <{\bar \psi }\psi> =\0{g_s ^2
}{m_s ^2 }\rho
_s\label{eq1},\eea 
where the scalar baryon density $\rho _s$ is defined as 
\bea
\rho _s =\0{\gamma }{(2 \pi )^3 }\int d^3 {\bf k} \0{M^*_N }{\om } \[n_N(\mu
^*,T)+{\bar n}_N(\mu
^*,T)\],
\eea
with the nucleon degenerate factor $\gamma =4$.

The effective chemical potential $\mu ^* $ is introduced by
\bea
\mu ^* =\mu -g^2_v \rho_N /m^2_v\label{eq2},\eea
with the baryon density
\bea
\rho _N =\0{\gamma}{(2\pi )^3 }\int d^3 {\bf k} \0{M^*_N}{\om }
[n_N(\mu 
^*,T)-{\bar n}_N(\mu
^*,T)].
\eea

One can solve the coupled equations (\ref{eq1}) and (\ref{eq2}) numerically. 
With the parameters 
$g_s ^2=109.626$, $g_v^2
=190.431$, $m_s  =520~MeV$, $M_N=938~ MeV$, and $m_v =783~ MeV$ according to 
the bulk binding energy
and normal density of nuclear matter at saturation,  the
effective nucleon mass and chemical potential with MFT are
displayed in Fig.\ref{figmass}.
\FIGURE{
\psfig{file=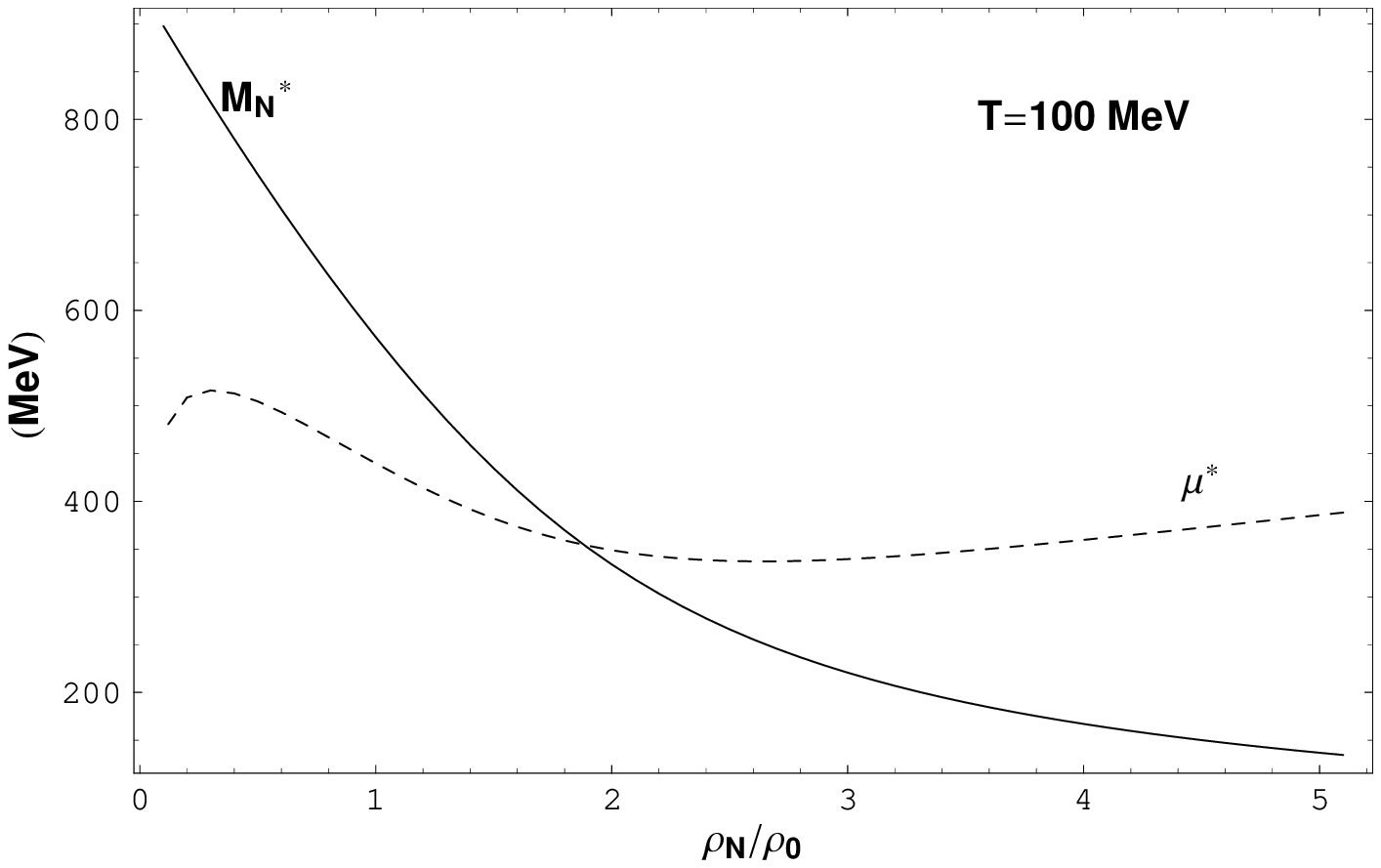,width=6.5cm,angle=-0}\hfill
\psfig{file=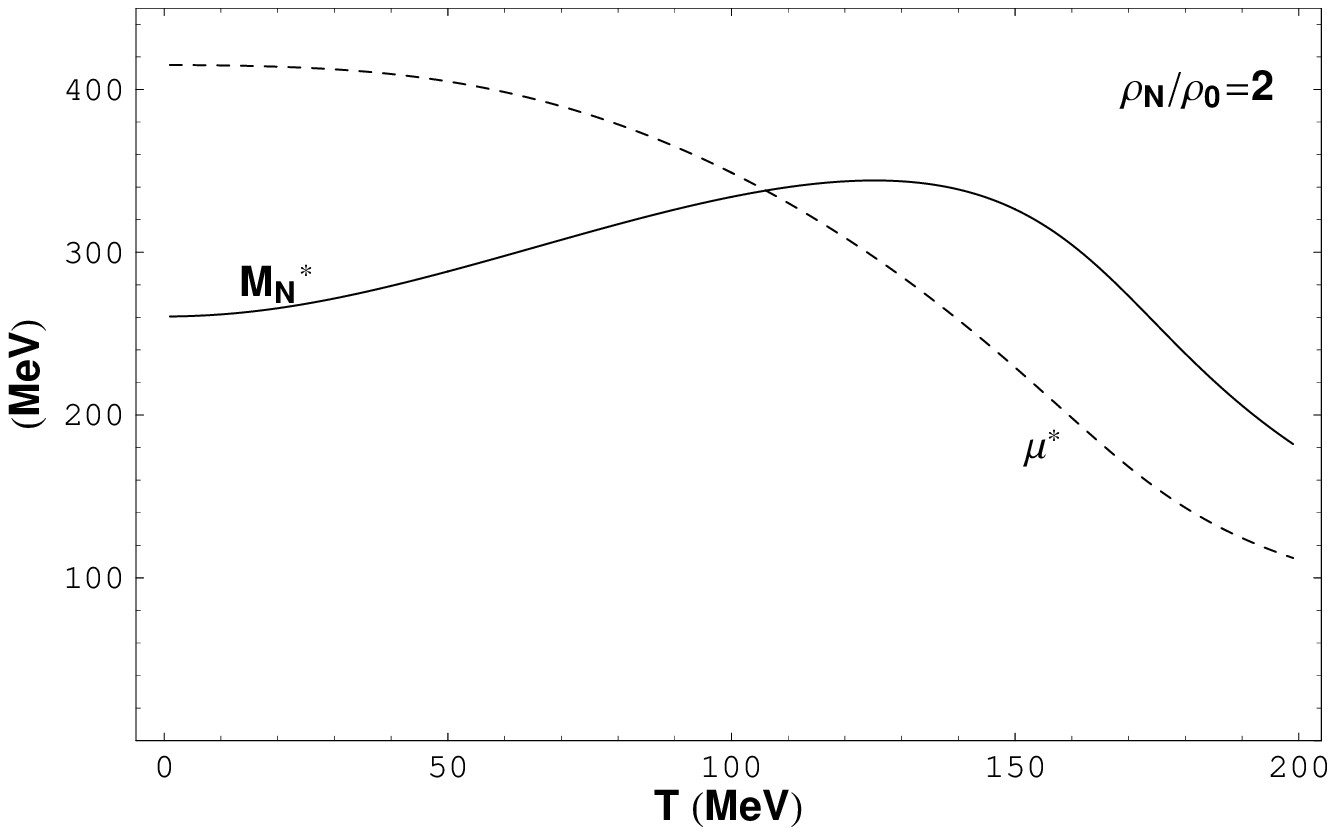,width=6.5cm,angle=-0} 
\caption{\small 
Effective nucleon mass $M_N^*$ and chemical potential $\mu ^*$ 
determined by QHD-I as functions of scaled  density (left panel ) and
temperature (right panel ) with MFT. The normal nuclear density is chosen to be 
$\rho _0 =0.16 ~fm^{-3}.$
}\label{figmass}
}

The longitudinal spectral functions (\ref{ten})
 in medium are drawn in
Figs.\ref{fig2} and \ref{fig3} for different hadronic
environment. The parameters are chosen to be $m_\pi =139.6 ~MeV$,
$m_\rho =770~ MeV$, $g_\rho^2 = 4 \pi * 2.91$, $\grnns =6.91$,
 $\kr =6.1$\refr{physrep}. 
It should be noted that, different from $\omega $, the tensor coupling is
very important for the interaction of $\rho$ with nucleons. Here we use the coupling constants $\grnn$ and $\kr $
determined from the fitting of the nucleon-nucleon interaction data done by
the Bonn group. 
As shown in Fig.\ref{fig2},
when nucleon-loop is not taken into account, there is only
temperature effect and  the $\rho $
meson spectral function changes weakly. With the increase of temperature, the
spectral function becomes a little bit wider and is shifted towards the
high invariant mass region slightly. This behavior agrees with the result of Ref.\cite{kapusta}.
\FIGURE{\psfig{file=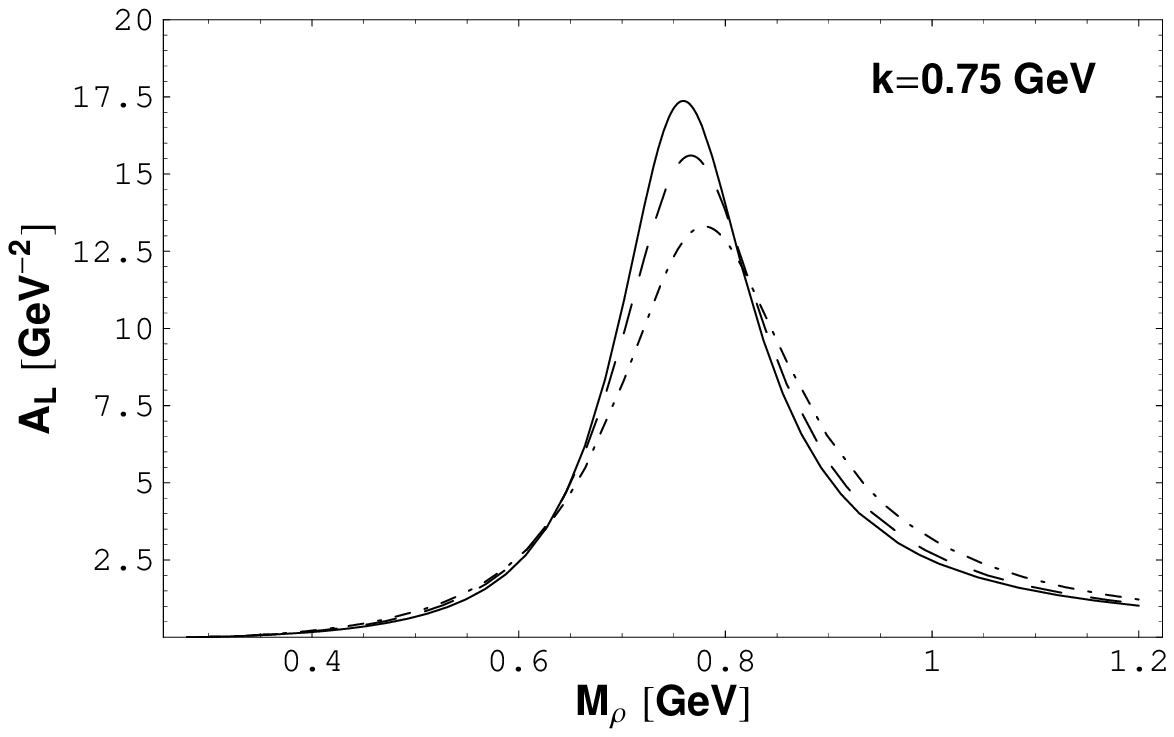,width=7.0cm,angle=-0}
\caption{\small The $\rho$ meson spectral function without considering the nucleon-loop at
momentum $k =0.75~GeV$. 
The solid line represents the vacuum situation, while
dashed and dot-dashed lines correspond to temperature
$T= 0.1$ and $0.15~ GeV$, respectively. }\label{fig2}
}

The spectral function including the nucleon-loop contribution is shown in
Fig.\ref{fig3}. The solid lines are still the calculation in vacuum, but
the dashed and dot-dashed lines contain the density  and
finite temperature effects. The dotted line represents the result of pure
density effect ($T=0$, $\rho _B\neq 0$). The density effect on $\rho $ spectral
function is obviously important by comparing Fig.\ref{fig2} with Fig.\ref{fig3}.
 The nucleon excitation shifts the spectral function towards low invariant mass
region. As indicated in the left panel of Fig.\ref{fig3}, in the case of low
density, $\rho
_N/\rho _0=0.1$,
the density effect is already remarkable compared with the pure temperature effect
shown in Fig.\ref{fig2}. 
The peak position of spectral function is
shifted from $0.77~ GeV $ in vacuum to about $0.63 ~GeV$ in medium.

The second characteristic of nucleon loop effect is the sharping of $\rho $
spectral function. Even for low density situation, see the left panel of  Fig.\ref{fig3}
with $\rho 
_N/\rho _0=0.1$, the
$\rho$ meson width is already reduced to about half of that in vacuum.
When density is high enough, see the right panel of Fig.\ref{fig3} with
$\rho 
_N/\rho _0=1.5$, the spectral function approaches to a $\delta $-like function.
This is mainly a consequence of the reduced two-pion phase space due to the $\rho $
mass decrease in medium.
\FIGURE{
\epsfig{file=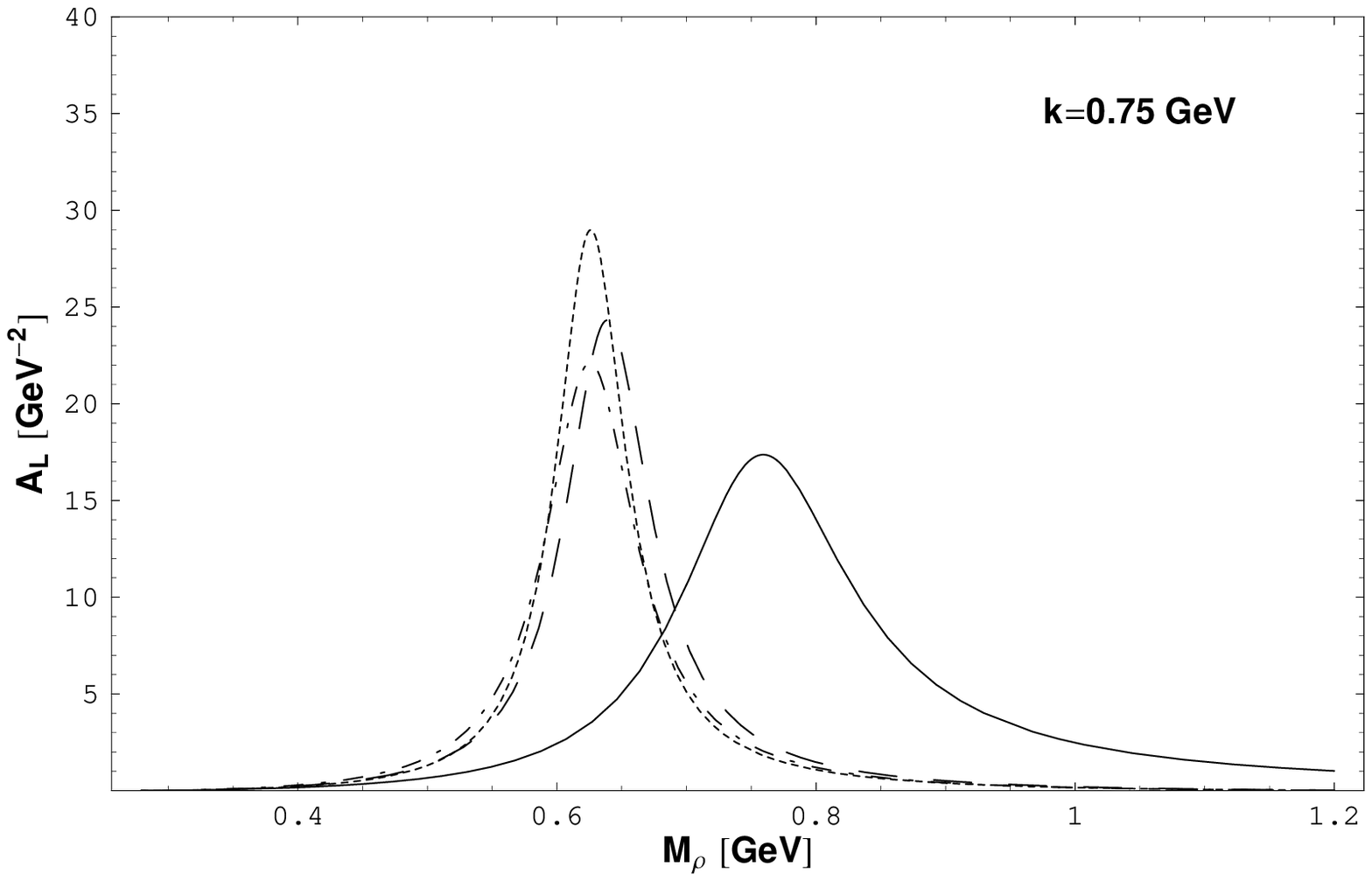,width=6.5cm }\hfill
\epsfig{file=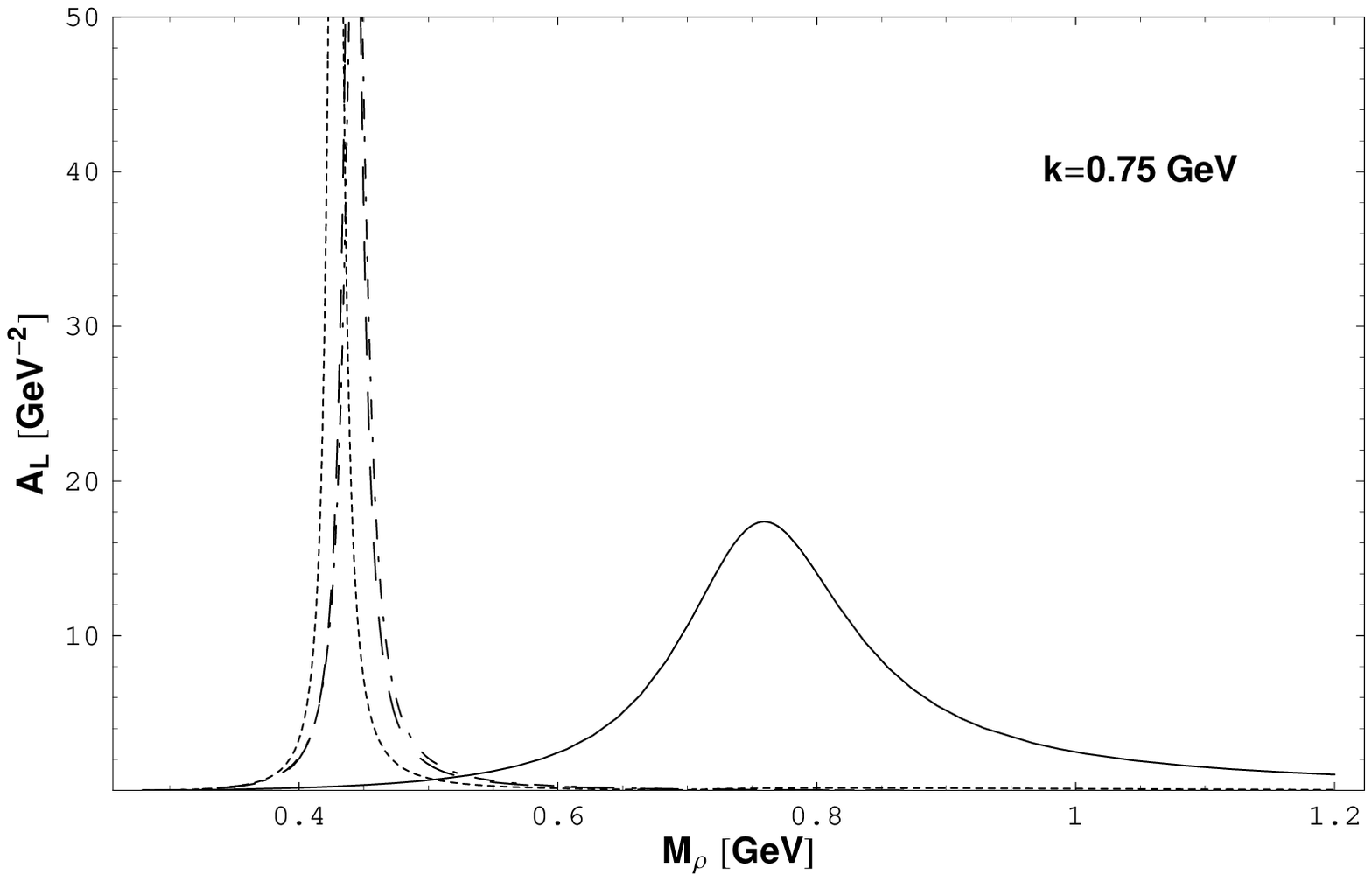,width=6.5cm}
 \caption{\small
Spectral function including the nucleon-loop contribution at momentum
$k=0.75~ GeV$.
Line style is similar to Fig.\ref{fig2} except that the dotted lines represent the
$T=0$ situation. 
Left panel: ${\rho _N}/{\rho _0} =0.1 $; Right panel:
${\rho _N}/{\rho _0} =1.5$.
}\label{fig3}}
\FIGURE{
\psfig{file=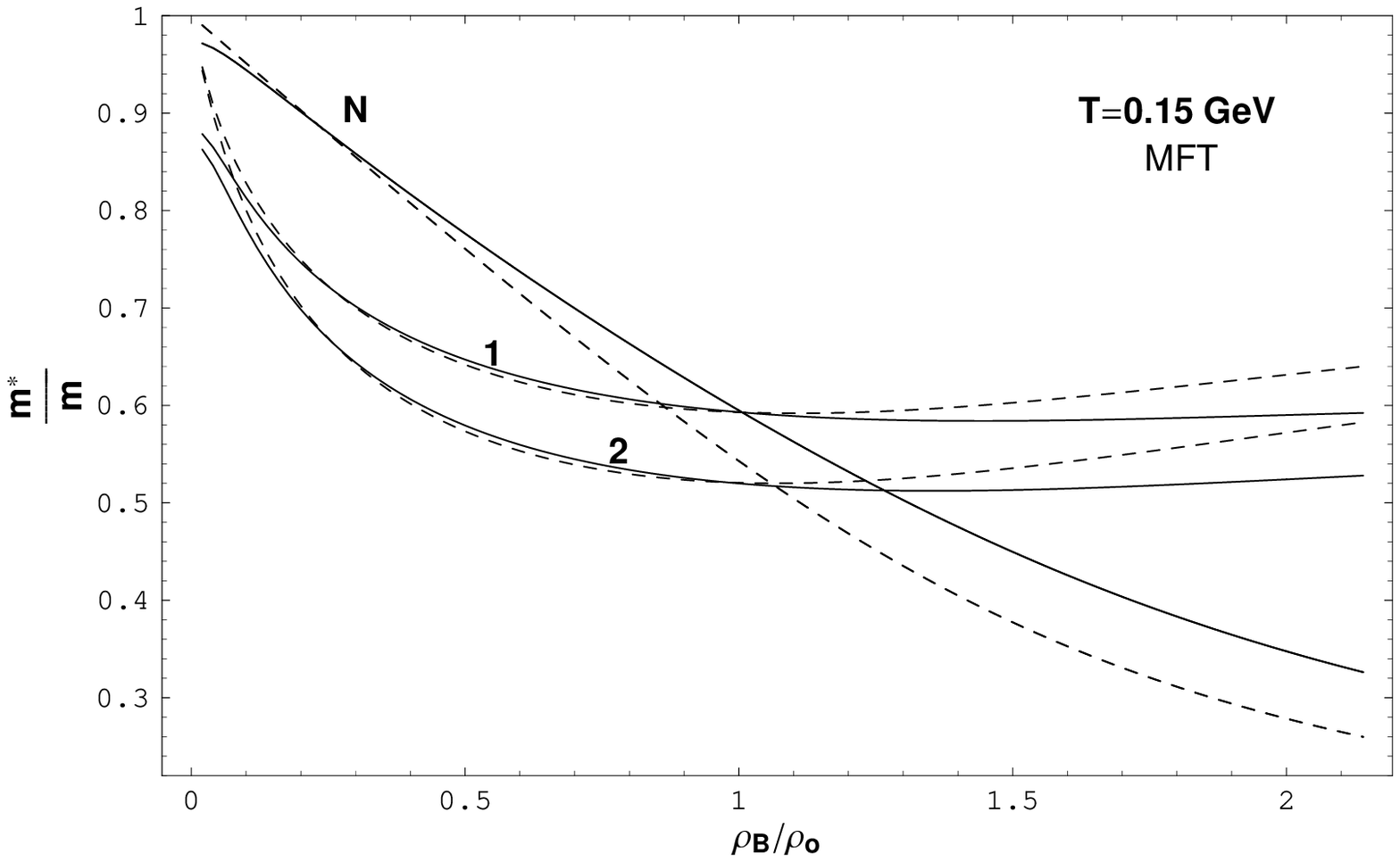,width=7.0cm,angle=-0}\hfill
\psfig{file=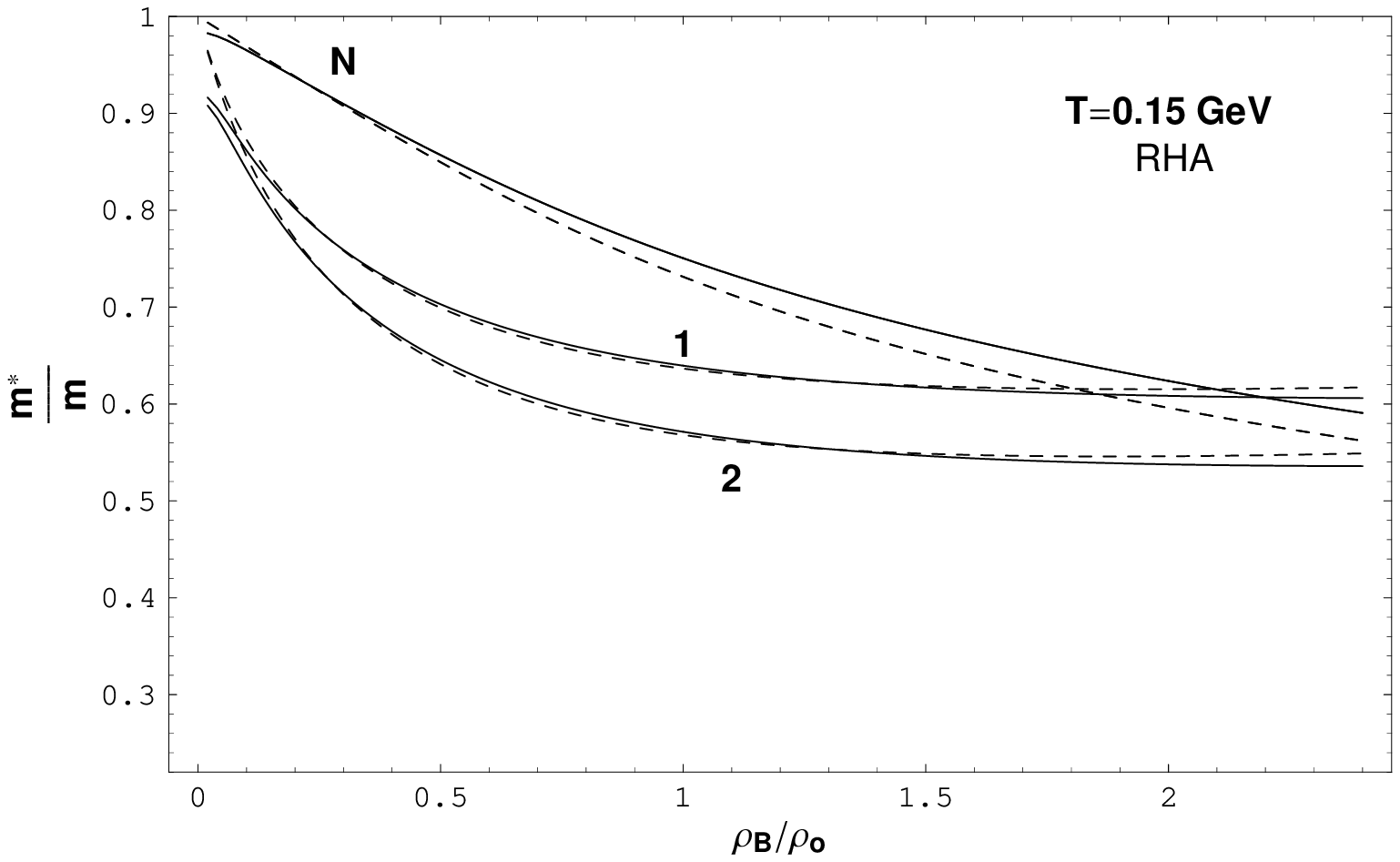,width=7.0cm,angle=-0} 
\caption{\small
Effective $\rho $ meson mass vs scaled baryon density for $T= 0.15~ GeV$
with MFT (left panel ) and RHA (right panel). For comparison, the effective nucleon mass line is also drawn out.
The dashed lines correspond to $T=0$ situation, while the solid lines to
those in the hot medium. The symbol ``1" represents the case considering form
factor, while ``2" without the form factor contribution. }\label{fig4}}

Another significance of the nucleon loop is its approximate saturation at high density.
At $\rho _N/\rho_0=1.5$, the two
lines with temperature $T=0.1 ~GeV $ and $0.15 ~GeV$ almost coincide with
each other. When density goes up further, for instance taking $\rho _N/\rho_0=2$, the
spectral function is almost the same as the corresponding line in the right
panel of Fig.\ref{fig3} for $\rho _N/\rho_0=1.5$.  This saturation tendency can be
 seen from the effective mass $m^*_\rho $ determined by the pole position of
its propagator in medium
\cite{shiomi}
\bea
k^2_0-m_\rho^2-\lim_{|\vk|\rightarrow 0}\Pi^{L(T)}(k)=0,
\eea
as indicated by Fig.\ref{fig4}.
For comparison,
we also give the effective nucleon mass $M_N^*$. By taking MFT, $m^*_\rho$ and 
$M_N^*$ are displayed in the left panel
of Fig. \ref{fig4}.  When density is higher than $\rho _0$, 
$m^*_\rho$ approaches its minimum value and then increases with the increase of
density a little bit. The result of $m^*_\rho $ with the nucleon
propagator obtained with the relativistic
Hartree approximation (RHA) by including vacuum fluctuation contribution instead of MFT 
is quite similar except the numerical difference, as
indicated in the right panel of Fig.\ref{fig4}.
Vector meson masses will decrease with the increase of density for a
model including the polarization of nucleon-antinucleon with a decreasing effective nucleon
mass\refr{shiomi,serot}. The numerical calculation indicates that the Dirac
sea contribution dominates.

It should be noted that the interactions between mesons and nucleons
are point-like in QHD. Considering the fact that mesons and nucleon are
composite, a phenomenological monopole form factor is multiplied at each meson-N
vertex\refr{physrep}
\bea
F_\alpha (k^2) = \0{\Lambda _\alpha ^2-m_\alpha ^2}{\Lambda ^2_\alpha -k_\mu
^2},
\eea
where $k_\mu$ is the four-momentum transfer, $m_\alpha $ the mass of the
exchanged meson and $\Lambda _\alpha $ the relevant cut-off mass. For $\rho
NN$ vertex, $\Lambda _\rho $ is usually taken as $\Lambda _\rho =1.5~GeV$.

The inclusion of form factor attenuates the behavior of effective
mass $m_\rho ^*$ significantly, which makes the magnitude of mass shift in medium smaller than that without the form factor. 
For analyzing the influence of form factor on the
effective mass $m^*_\rho $, the two situations with/without form factor
contribution have been drawn out in Fig.\ref{fig4}.
The displayed spectral function in Fig. \ref{fig3} has included the form factor
contribution.

Our result of the spectral function is different from previous results such as in
\refrs{rapp,rapp2}. From Figs.\ref{fig3} and \ref{fig4}, one can see that effects of the decreasing
effective nucleon mass 
on $\rho $ spectral function are important but
they are not considered in
\refrs{rapp,rapp2}. On the other
hand, by focusing on the effects of decreasing
nucleon mass on QHD-I level, any other decay channels which can contribute to
the imaginary as well as the real parts of the self-energy are not
included in our calculation and also cause the difference.

In summary, we discussed the density and temperature effects on $\rho$ meson
spectral function with the effective Lagrangian. Different from the pure temperature effect of the pion
loop which changes $\rho$
meson mass only a little and shifts the spectral function towards high
invariant mass region slightly, the nucleon-loop changes the $\rho $ meson
properties differently and significantly:
1) The spectral function is shifted towards low mass region remarkably
even in the case of low density; 
2) Together with the shift towards low
mass region, the spectral function becomes very sharp due to Dirac sea
polarization with the decreasing nucleon mass in medium; 
The density effect on the spectral function approaches to
saturation when the density increases. 3) The influence of a
phenomenological vertex form factor may prevent the effective $\rho $ mass decreasing
to some extent within hadronic level. 
Although the behavior of $m_\rho^*$ in low density region is expected to be not so
stiff, our result indicates that the effects of decreasing nucleon mass may be
remarkable and should be taken into account in discussing the in-medium
$\rho $ property. The relation between our result and those from QCD sum
rules esp. in the low density region is not yet well understood and needs to be clarified further.
\acknowledgments{The work was supported in part by the NSFC and the 973 Project.}

\end{document}